\documentclass[aps,prl,reprint,superscriptaddress,showpacs,citeautoscript,nopacs,twocolumn]{revtex4}

\setcitestyle{super}
\usepackage{graphicx}
\usepackage{color}
\usepackage{amsfonts}

\begin{document}
\title{Quasiparticle dynamics and spin-orbital texture of the SrTiO$_3$ two-dimensional electron gas}

\author{P.~D.~C.~King}
\email{philip.king@st-andrews.ac.uk}
\affiliation{Kavli Institute at Cornell for Nanoscale Science, Ithaca, New York 14853, USA}
\affiliation{Laboratory of Atomic and Solid State Physics, Department of Physics, Cornell University, Ithaca, New York 14853, USA}
\affiliation{SUPA, School of Physics and Astronomy, University of St. Andrews, St. Andrews, Fife KY16 9SS, United Kingdom}

\author{S.~McKeown Walker}
\affiliation{D{\'e}partement de Physique de la Mati{\`e}re Condens{\'e}e, Universit{\'e} de Gen{\`e}ve, 24 Quai Ernest-Ansermet, 1211 Gen{\`e}ve 4, Switzerland}

\author{A.~Tamai}
\affiliation{D{\'e}partement de Physique de la Mati{\`e}re Condens{\'e}e, Universit{\'e} de Gen{\`e}ve, 24 Quai Ernest-Ansermet, 1211 Gen{\`e}ve 4, Switzerland}

\author{A.~de~la~Torre}
\affiliation{D{\'e}partement de Physique de la Mati{\`e}re Condens{\'e}e, Universit{\'e} de Gen{\`e}ve, 24 Quai Ernest-Ansermet, 1211 Gen{\`e}ve 4, Switzerland}

\author{T.~Eknapakul}
\author{P. Buaphet}
\affiliation{School of Physics and NANOTEC-SUT Center of Excellence on Advanced Functional Nanomaterials, Suranaree University of Technology, Nakhon Ratchasima 30000, Thailand}

\author{S.-K.~Mo}
\affiliation{Advanced Light Source, Lawrence Berkeley National Lab, Berkeley, CA 94720, USA}

\author{W. Meevasana}
\affiliation{School of Physics and NANOTEC-SUT Center of Excellence on Advanced Functional Nanomaterials, Suranaree University of Technology, Nakhon Ratchasima 30000, Thailand}

\author{M.~S.~Bahramy}
\affiliation{Quantum-Phase Electronics Center and Department of Applied Physics, The University of Tokyo, Tokyo 113-8656, Japan}
\affiliation{RIKEN center for Emergent Matter Science (CEMS), Wako 351-0198, Japan}

\author{F.~Baumberger}
\email{Felix.Baumberger@unige.ch}
\affiliation{D{\'e}partement de Physique de la Mati{\`e}re Condens{\'e}e, Universit{\'e} de Gen{\`e}ve, 24 Quai Ernest-Ansermet, 1211 Gen{\`e}ve 4, Switzerland}
\affiliation{Swiss Light Source, Paul Scherrer Institut, CH-5232 Villigen PSI, Switzerland}
\affiliation{SUPA, School of Physics and Astronomy, University of St. Andrews, St. Andrews, Fife KY16 9SS, United Kingdom}

\date{\today}
%

%\pacs{}

\begin{abstract}\noindent Two-dimensional electron gases (2DEGs) in SrTiO$_3$ have become model systems for engineering emergent behaviour in complex transition metal oxides. Understanding the collective interactions that enable this, however, has thus far proved elusive. Here we demonstrate that angle-resolved photoemission can directly image the quasiparticle dynamics of the $d$-electron subband ladder of this complex-oxide 2DEG. Combined with realistic tight-binding supercell calculations, we uncover how quantum confinement and inversion symmetry breaking collectively tune the delicate interplay of charge, spin, orbital, and lattice degrees of freedom in this system. We reveal how they lead to pronounced orbital ordering, mediate an orbitally-enhanced Rashba splitting with complex subband-dependent spin-orbital textures and markedly change the character of electron-phonon coupling, co-operatively shaping the low-energy electronic structure of the 2DEG. Our results allow for a unified understanding of spectroscopic and transport measurements across different classes of SrTiO$_3$-based 2DEGs, and yield new microscopic insights on their functional properties.\end{abstract}

\maketitle

\begin{figure*}[!t]
\begin{center}
\includegraphics[width=\textwidth]{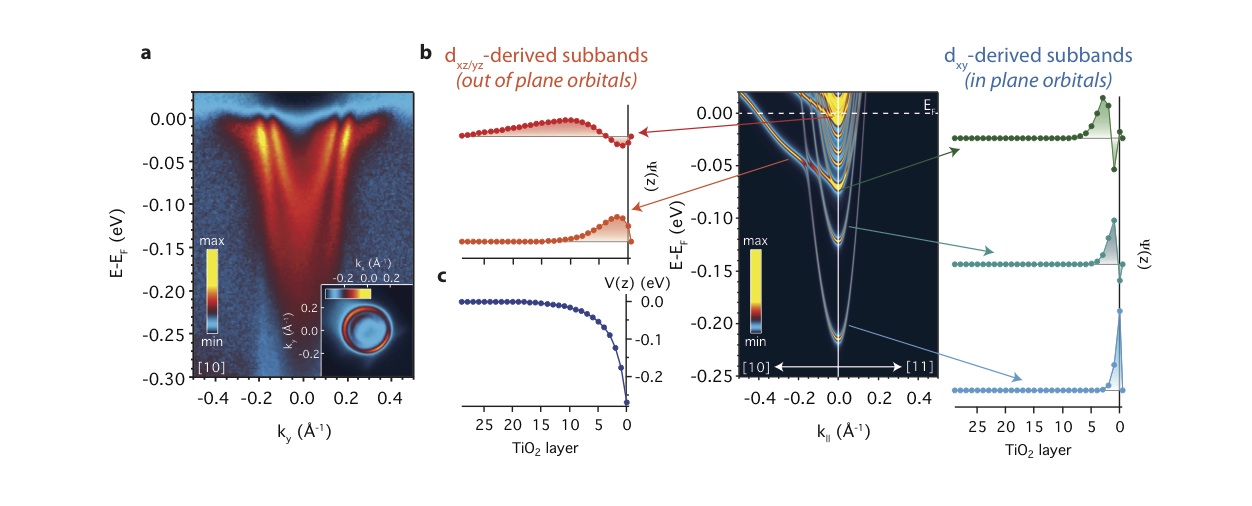}
\caption{ \label{f:overview} {\bf Orbital ordering of a two-dimensional electron gas in SrTiO$_3$.} (a) $E$ vs.\ $k$ dispersion from ARPES ($h\nu=50$~eV, measured along the $[10]$ direction), revealing a multi-orbital subband structure comprising co-existing ladders of light and massive $d$-electron subband states. The respective circular and faint elliptical Fermi surface pockets (measured with $h\nu=51$~eV and the polarization along $[11]$) are shown in the inset. For the dispersion plot, a normalisation (division by the average MDC) has been applied to better reveal the massive band, as shown in Supplementary Fig.~1. (b) This electronic structure is well described by a self-consistent tight-binding supercell calculation. The spatial dependence of the subband wavefunctions along the confinement direction, $\Psi(z)$, reveal a pronounced real-space orbital ordering, a direct consequence of near-surface band bending (c).}
\end{center}
\end{figure*}
	The ubiquitous perovskite oxide SrTiO$_3$, a wide-gap band insulator, hosts varied bulk properties including quantum paraelectricity, dilute doping-induced superconductivity and high thermoelectric coefficients. These reflect a subtle competition between interactions of the underlying quantum many-body system. Intriguingly, thermodynamic and transport measurements~\cite{Mannhart:Science:327(2010)1607--1611,Hwang:NatureMater.:11(2012)103--113,Reyren:Science:317(2007)1196--1199,Caviglia:Nature:456(2008)624--627,Bert:NaturePhys.:7(2011)767--771,Li:NaturePhys.:7(2011)762--766} indicate that the balance of these interactions can be tuned to engineer striking emergent properties when quantum confinement and doping are combined to create a two-dimensional electron gas (2DEG).~\cite{Mannhart:Science:327(2010)1607--1611,Hwang:NatureMater.:11(2012)103--113,Meevasana:NatureMater.:10(2011)114--118}  A diverse and attractive array of properties have been uncovered to date in this system, including gate-tuned superconductivity,~\cite{Reyren:Science:317(2007)1196--1199,Caviglia:Nature:456(2008)624--627,Ueno:NatureMater.:7(2008)855--858} its coexistence with ferromagnetism,~\cite{Bert:NaturePhys.:7(2011)767--771,Li:NaturePhys.:7(2011)762--766} and enhanced Seebeck coefficients,~\cite{Ohta:NatMater:6(2007)129--134} establishing SrTiO$_3$ based 2DEGs as a model platform for use in future multifunctional electronic devices.~\cite{Mannhart:Science:327(2010)1607--1611}

They are most commonly realised at a polar interface to another band insulator LaAlO$_3$, creating a narrow conducting channel that resides solely within the SrTiO$_3$.~\cite{Ohtomo:Nature:427(2004)423--426} Similar 2DEGs can also be created by interfacing SrTiO$_3$ to a wide array of other band or Mott insulators including NdAlO$_3$,~\cite{Ariando2012} LaTiO$_3$,~\cite{Ohtomo:Nature:419(2002)378} and GdTiO$_3$,~\cite{Moetakef:Appl.Phys.Lett.:99(2011)232116} by chemical doping of electrons into narrow SrTiO$_3$ channels,~\cite{Kozuka:Nature:462(2009)487--490,Choi:NanoLett.:12(2012)4590--4594} analogous to $\delta$-doping of semiconductors such as Si, and by field-effect doping in a transistor-style configuration.~\cite{Ueno:NatureMater.:7(2008)855--858} Moreover, the recent discovery of a 2DEG formed at the free surface of a bulk SrTiO$_3$ crystal opens new avenues for its advanced spectroscopic investigation.~\cite{Meevasana:NatureMater.:10(2011)114--118,Santander-Syro:Nature:469(2011)189--193} 

Exploiting this, here we present unified angle-resolved photoemission (ARPES) measurements and tight-binding supercell calculations revealing new richness of the electronic structure of this model oxide 2DEG.  We show how a pronounced orbital ordering mediates an unconventional spin splitting, giving rise to strongly anisotropic and subband-dependent canted spin-orbital textures. The orbitally enhanced Rashba effect explains the pronounced spin splittings previously inferred from magnetotransport in this system, while simultaneously revealing a breakdown of the conventional picture used to describe these. We uncover how this complex ladder of subband states are further renormalized by many-body interactions. This reconciles previous discrepancies between effective masses estimated from ARPES and quantum oscillations, unifying the properties of surface and interface SrTiO$_3$ 2DEGs, and reveals a strikingly different nature of electron-phonon coupling compared to bulk SrTiO$_3$. 

\newpage
{\large\noindent{\bf Results}}\\
\noindent{\bf Orbital ordering.} Figure~\ref{f:overview} summarises the generic electronic structure of SrTiO$_3$ 2DEGs, as revealed by ARPES from a SrTiO$_3$(100) surface with saturated band bending.~\cite{Meevasana:NatureMater.:10(2011)114--118} Consistent with previous reports from both surface and interface 2DEGs,~\cite{Meevasana:NatureMater.:10(2011)114--118,Santander-Syro:Nature:469(2011)189--193,D'Angelo:Phys.Rev.Lett.:108(2012)116802,Berner:Phys.Rev.Lett.:110(2013)247601,Plumb:arXiv:1302.0708:(2013)} we find a broad bandwidth that extends up to $\approx250$~meV below the Fermi level. Here, we can resolve a ladder of at least three light subband states that contribute concentric circular Fermi surface sheets, co-existing with just a single heavy electron band ($m^*=14\pm3m_\mathrm{e}$) that has a much shallower binding energy of less than $50$~meV and gives rise to elliptical Fermi surfaces oriented along $\langle10\rangle$. From this Fermi surface topology together with the polarization dependence of our measured intensities (Supplementary Fig.~2), we assign not only the lowest,~\cite{Santander-Syro:Nature:469(2011)189--193} but rather the whole ladder of observed light states as having dominantly  $d_{xy}$ orbital character, while the heavy states derive from $d_{xz/yz}$ orbitals. This immediately indicates a strong breaking of the $t_{2g}$ orbital degeneracy that is present in the bulk electronic structure of SrTiO$_3$,~\cite{Salluzzo:Phys.Rev.Lett.:102(2009)166804} driving a pronounced orbital ordering with a polarization $P=\frac{n(d_{xy})-n(d_{xz/yz})}{n(d_{xy})+n(d_{xz/yz})}$, which exceeds 30\%, a lower limit derived from our experimentally-resolved Fermi surface areas.
 
This is a direct consequence of the real-space anisotropy of the orbital wavefunctions combined with inversion symmetry breaking by the electrostatic potential that defines the 2DEG by creating a steep asymmetric quantum well along the $z$-direction (Fig.~\ref{f:overview}(c)). As shown by our self-consistent tight-binding supercell calculations (Fig.~\ref{f:overview}(b), see Methods), the resulting quantized subbands that derive from planar $d_{xy}$ orbitals have wavefunctions reminiscent of the envelope functions of a semiconductor quantum well, except that in SrTiO$_3$ they are much more localized in real space, almost to within a single unit cell for the lowest subband state. In contrast, the potential variation acts as a much weaker perturbation on the out-of-plane $d_{xz/yz}$ orbitals, which have much larger hopping amplitudes along the $z$-direction. The resulting subbands sit close to the top of the potential well, leading to wavefunctions that penetrate much deeper into the bulk. This disparate spatial extent of the subband states is consistent with their relative spectral weight in our surface-sensitive ARPES measurements.
\begin{figure*}[!t]
\begin{center} 
\includegraphics[width=\textwidth]{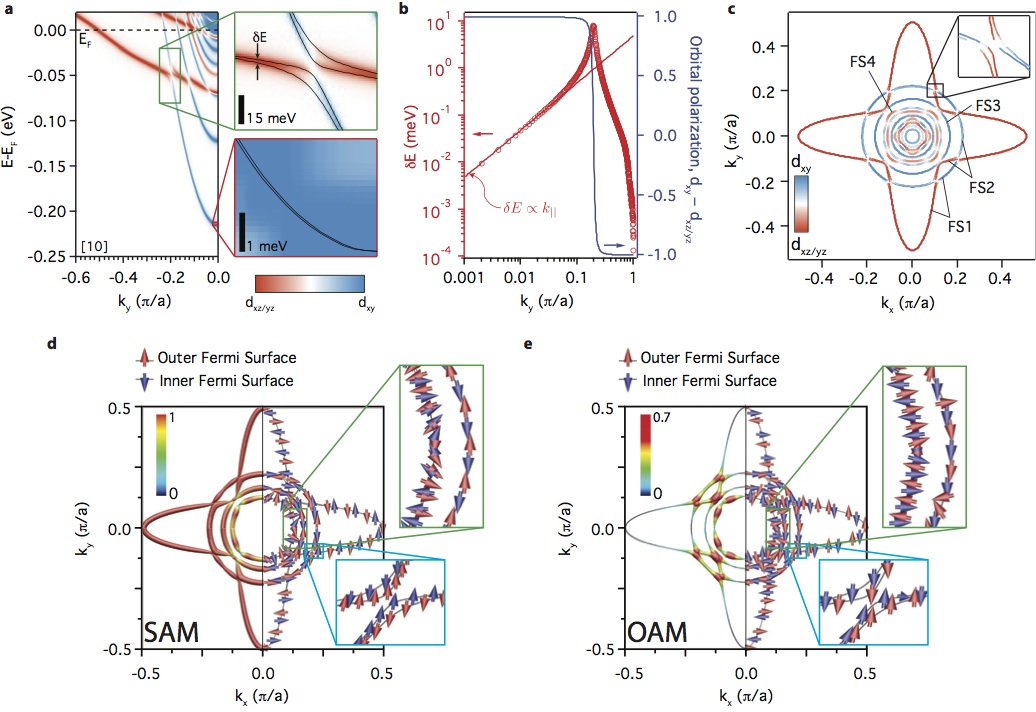}
\caption{ \label{f:Rashba} {\bf Orbitally-enhanced spin splitting.} (a) Orbitally-resolved electronic structure of the 2DEG along the $[10]$ direction. Magnified views reveal a weak Rashba-type spin splitting around the band bottom, which becomes enhanced by approximately an order of magnitude at the crossings of the light and heavy subband states where the orbital character becomes strongly mixed, as quantified in (b) for the lowest subbands. (c) The calculated Fermi surface shows how similar orbital mixing and pronounced spin splittings occur at the crossings of circular $d_{xy}$- and elliptical $d_{xz/yz}$-derived Fermi surface sheets. This gives rise to an exotic anisotropic and subband-specific coupled spin-orbital texture of the 2DEG, as evident from the magnitude (left) and direction (right) of the (d) spin, SAM, and (e) orbital, OAM, angular momenta of the four largest Fermi surface sheets. The magnitude of the SAM (OAM) is represented on a false colour scale in units of $\hbar/2$ ($\hbar$), respectively, and is shown for additional Fermi surfaces in Supplementary Fig.~4.}
\end{center}
\end{figure*}
\\
\
\\
\noindent{\bf Unconventional Rashba spin splitting.} The same breaking of inversion symmetry that drives this orbital ordering can additionally lift the spin degeneracy through a Bychkov-Rashba-type spin-orbit interaction.~\cite{Bychkov:JETPLett.:39(1984)78,Grundler:Phys.Rev.Lett._84:6074--6077(2000),King:Phys.Rev.Lett.:107(2011)096802} Focussing near the band bottom of the lowest $d_{xy}$ band (Fig.~\ref{f:Rashba}(a,b)), we indeed find a small characteristic splitting between the calculated energy of spin-up and spin-down states, $\delta{E}=2\alpha{}k_{||}$, with a Rashba parameter, $\alpha=0.003$~eV\AA{}. The non-negligible splitting found here, despite the modest spin-orbit interaction in 3$d$ transition metals, is indicative of the very strong electric field gradient of the confining potential. From our self-consistent band bending potential calculation (Fig.~\ref{f:overview}(c)), we estimate that this exceeds $2\times10^8$~Vm$^{-1}$ within the first 2 unit cells where the lowest subband state is confined. For the more delocalized second $d_{xy}$ subband, whose wavefunction extends into regions of shallower band bending, we find a slightly smaller Rashba parameter $\alpha=0.0014$~eV\AA, confirming that the strength of this spin splitting is controlled by the confining electric field. This should therefore be directly tuneable by electrical gating, suggesting a potential route towards spintronic control in oxides. 

Unlike typical Rashba systems such as the Au$(111)$ surface, however, here the interplay between orbital ordering and spin-orbit coupling leads to a significantly richer spin structure of the 2DEG states. Close to the crossings of the light $d_{xy}$ and heavy $d_{xz/yz}$ subbands, the spin splitting increases by approximately an order of magnitude, concomitant with a strong mixing of their orbital character (Fig.~\ref{f:Rashba}(a-c)). This rationalises an increased spin splitting reported from transport when the $d_{xz/yz}$ subbands become populated in electrically-gated SrTiO$_3$/LaAlO$_3$ interface 2DEGs.~\cite{Fete:Phys.Rev.B:86(2012)201105} Moreover, the crossover from $k$-linear to strongly enhanced spin splitting that we find here readily explains the approximately $k^3$ dependence of the splitting that has been reported.~\cite{Nakamura:Phys.Rev.Lett.:108(2012)206601} Our layer-projected calculations indicate that the subband wavefunctions become more delocalized in the $z$-direction close to these band crossings, a natural consequence of the stronger overlap of neighbouring $d_{xz/yz}$ orbitals along $z$ (Supplementary Fig.~3). This delocalization would naively be expected to reduce the strength of the Rashba effect. In contrast, its significant enhancement here points to a dominant role of inter-orbital hopping in driving such surprisingly-large spin splittings. Similar enhancements have recently also been observed in other calculations, mainly based on model 3-band Hamiltonians,~\cite{Zhong:Phys.Rev.B:87(2013)161102,Khalsa:arXiv:1301.2784:(2013),Kim2013} which are qualitatively entirely consistent with our results. Our calculations demonstrate how this is a direct consequence of orbital ordering in the real experimentally-confirmed multi-subband structure of the SrTiO$_3$ 2DEG. 

Moreover, as shown in Fig.~\ref{f:Rashba}(d,e), they reveal an exotic coupled spin-orbital texture of the resulting 2DEG Fermi surfaces. While an approximately perpendicular spin-momentum locking ensures tangential spin winding around the circular $d_{xy}$ sections of Fermi surface, it leads to spins aligned almost perpendicular to the Fermi surface for large sections of the extremely anisotropic  $d_{xz}$ ($d_{yz}$) sheets. Around the crossings of $d_{xy}$ and $d_{xz/yz}$ states, the spins of neighbouring Fermi surfaces align (anti-)parallel with a $\left|\downarrow\right\rangle\!\left|\uparrow\right\rangle\!\left|\uparrow\right\rangle\!\left|\downarrow\right\rangle$ ordering, ensuring maximal hybridization gaps are opened. At the same time, rather than being quenched to zero as might have naively been expected, we find a finite orbital angular momentum (OAM, $\bf{L}$) emerges. This is relatively small ($\lesssim0.05\hbar$) for the isolated $d_{xy}$ and $d_{xz/yz}$ sections of Fermi surface (Fig.~\ref{f:Rashba}(e) and Supplementary Fig.~3), but grows as large as $0.7\hbar$ around the band crossings. Moreover, we find that the OAM is oriented either parallel or antiparallel to the corresponding spin angular momentum (SAM, $\bf{S}$), and so this increase in OAM maximally enhances $\bf{L}\cdot\bf{S}$, by a factor of $\sim\!14$, at and in the vicinity of the hybridization gaps, comparable to the corresponding increase in spin splitting (Fig.~\ref{f:Rashba}(b)).

This therefore provides a natural basis to understand the large spin splittings inferred from magnetotransport,~\cite{Caviglia:Phys.Rev.Lett.:104(2010)126803,BenShalom:Phys.Rev.Lett.:104(2010)126802,Joshua:NatureCommun.:3(2012)1129--} despite the small atomic spin-orbit interaction of SrTiO$_3$, in terms of an orbital Rashba effect.~\cite{orb_rashba1,orb_rashba2} For isolated $d_{xy}$ sections of Fermi surface, we find that the OAM of the inner and outer spin-split branches have the same helicity, consistent with a weak spin-orbit coupling limit.~\cite{orb_rashba2} For the $d_{xz/yz}$ sections of Fermi surface, however, the inner and outer branches have opposite OAM, reflecting additional richness as compared to model systems such as noble metal surface states.~\cite{orb_rashba2} This causes mixed helicities of the OAM around the inner branch of each Fermi surface sheet, as compared to a complete $2\pi$ winding for the outer branches (see arrows in Supplementary Fig.~4). Importantly, we find that the dominant inter-band interactions cause the winding direction of both the OAM and SAM of the outer $d_{xy}$-derived Fermi surfaces to abruptly switch sign across the hybridization gaps. For several inner Fermi surfaces whose orbital character is strongly mixed, continuously evolving between $d_{xy}$- and $d_{xz/yz}$-like around the Fermi surface (Fig.~\ref{f:Rashba}(c)), this leads to strongly frustrated spin and orbital textures, rapidly canting from tangential to radial alignment as a function of Fermi surface angle (Fig.~\ref{f:Rashba}(d,e)). This is quite distinct from the functional form of conventional Rashba splitting~\cite{Bychkov:JETPLett.:39(1984)78} and provides strong constraints for the influence of spin-splitting on magnetism~\cite{Fischer:NewJ.Phys.:15(2013)023022} and superconductivity.~\cite{Michaeli:Phys.Rev.Lett.:108(2012)117003}
\begin{figure*}[!t]
\begin{center}
\includegraphics[width=\textwidth]{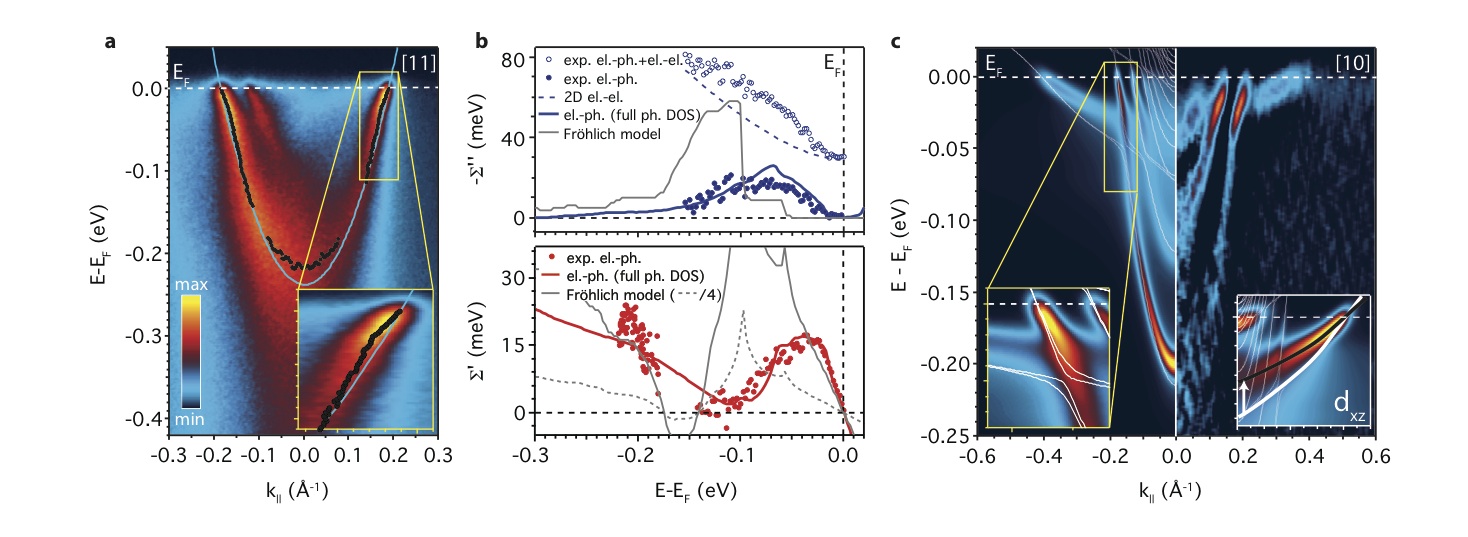}
\caption{ \label{f:e-ph} {\bf Quasiparticle dynamics of the subband states.} (a) ARPES measurements along the $[11]$ direction ($h\nu=51$~eV), together with the peak positions of fits to MDCs and EDCs (black dots) and a cosine `bare band' dispersion (blue line). The data reveal pronounced signatures of electron-phonon coupling. Particularly apparent is a low-energy kink in the dispersion at $\omega\approx30$~meV, shown magnified inset. (b) Real, $\Sigma'$, and imaginary, $\Sigma''$, parts of the extracted self-energy. The open blue symbols show $\Sigma''$ before subtraction of an electron-electron scattering contribution to the total measured imaginary part of the self-energy, approximated here by the expression $\Sigma_{\mathrm{e-e}}''(\omega)=\beta\omega^2[1+0.53 \ln(\omega/E_{\mathrm{F}})]$ for a 2D Fermi liquid (dashed line, $\beta=1.5$~eV$^{-1}$, $E_{\mathrm{F}}=0.25$~eV). The solid lines are calculated electron-phonon self energies with the Eliashberg function proportional to the full phonon density of states (red/blue) or for coupling to three longitudinal optical phonons (black), which dominate the interaction in lightly doped bulk SrTiO$_3$. (c, left) Calculated electron-phonon spectral function, $A(k,\omega)$, along the $[10]$ direction (see Methods). This reveals complete bandwidth renormalization of the heavy states (particularly apparent when $A(k,\omega)$ is projected only onto $d_{xz}$ orbitals as shown in the right inset), and kinks in the light $d_{xy}$ subband dispersions at their crossings with this renormalized heavy state (left inset), both in good agreement with our experimental measurements, shown in the right half using curvature analysis~\cite{Zhang:Rev.Sci.Instrum.:82(2011)043712} of the raw data. }
\end{center}
\end{figure*}
\\
\
\\
\noindent{\bf Many-body interactions.} In Figure~\ref{f:e-ph}, we further uncover a pronounced role of electron-phonon interactions on this complex hierarchy of electronic states. Unlike in bulk-doped SrTiO$_3$, where the Fermi energy is typically only a few meV and the electron-phonon interaction is thus non-retarded, the occupied widths of different subbands of the 2DEG range from almost zero up to values greater than the highest phonon frequency of $\approx100$~meV. This is an unusual situation, neither described by the adiabatic $(\hbar\omega_\mathrm{D}<<E_\mathrm{F})$ nor the anti-adiabatic $(\hbar\omega_\mathrm{D}>>E_\mathrm{F})$ approximation, and points to a complex influence of electron-phonon coupling in this system. We extract the corresponding self-energy, $\Sigma_{\mathrm{e-ph}}(\omega)=\Sigma'(\omega)+i\Sigma''(\omega)$, from our ARPES measurements of the lowest subband along the $[11]$ direction (see methods), where we resolve an isolated band all the way up to $E_\mathrm{F}$. The slope of $\Sigma'$ at the Fermi level yields an electron-phonon coupling strength of $\lambda=0.7(1)$, while its broad maximum between $\approx20 - 60$~meV is indicative of coupling to multiple modes. Indeed, the experimentally-determined self-energy is in excellent agreement with a calculation within Eliashberg theory that assumes a coupling function $\alpha^2 F(\omega)$ proportional to the entire phonon density of states associated with the motion of oxygen and Ti ions~\cite{Choudhury:Phys.Rev.B:77(2008)134111} and includes the realistic 2DEG electron density of states from our tight-binding calculation. 

Together with a moderate correlation-induced mass enhancement of $\approx1.4$ that we estimate from a Kramers-Kronig transform of a Fermi-liquid contribution to the imaginary part of the self-energy, our analysis suggests an overall mass enhancement arising from many-body interactions of $m^{*}/m_{\rm{band}}\approx2.1$, close to the values deduced for lightly-doped bulk SrTiO$_3$ from measurements of the electronic specific heat~\cite{Ambler:Phys.Rev.:148(1966)280--286} and optical spectroscopy~\cite{Mechelen:Phys.Rev.Lett.:100(2008)226403}. The nature of the electron-phonon coupling, however, is very different. In lightly-doped bulk SrTiO$_3$, it is dominated by the long-range coupling to longitudinal optical (LO) phonons as described by the Fr\"ohlich model.~\cite{Verbist:Ferroelectrics:130(1992)27--34, Mechelen:Phys.Rev.Lett.:100(2008)226403}  This model predicts much weaker coupling to low-energy modes than observed here, but a significantly stronger coupling to the highest LO phonon at 100~meV, as evident from a calculation employing coupling strengths from bulk SrTiO$_3$,~\cite{Verbist:Ferroelectrics:130(1992)27--34,Meevasana:NewJ.Phys.:12(2010)023004} which yield an electron-phonon self-energy in clear contrast to our experimental findings (Fig.~\ref{f:e-ph}(b)).

The electric field that confines the 2DEG is known to dramatically reduce its dielectric constant.~\cite{Copie:Phys.Rev.Lett.:102(2009)216804,Meevasana:NatureMater.:10(2011)114--118} This, together with higher carrier densities as compared to bulk SrTiO$_3$, will lead to shorter electronic screening lengths for the 2DEG, explaining the observed suppression of long-range coupling to LO modes. The enhanced coupling to low-energy phonons for the 2DEG instead leads to a pronounced kink in the dispersion of the $d_{xy}$ subbands at an energy around $30$~meV. We resolve these along both the $[10]$ and $[11]$ directions for the first two $d_{xy}$ subbands. Crucially, the resulting enhanced quasiparticle mass, which we estimate as $1.1(2)$~m$_e$ from our measured Fermi velocities, rectifies the discrepancy between the light masses around 0.6~m$_e$ reported in earlier ARPES studies of SrTiO$_3$ surface 2DEGs~\cite{Meevasana:NatureMater.:10(2011)114--118, Santander-Syro:Nature:469(2011)189--193} and recent quantum oscillation experiments that revealed effective masses typically around $1m_\mathrm{e}$.~\cite{Caviglia:Phys.Rev.Lett.:105(2010)236802,Kim:Phys.Rev.Lett.:107(2011)106801}

Intriguingly, along $[10]$ the kink energy coincides almost exactly with the crossing of the light $d_{xy}$ and heavy $d_{xz}$ subband states. This behaviour is well captured by our spectral function simulations calculated from our tight-binding bare dispersions and electron-phonon self-energy. These illustrate a very different effect of electron-phonon interactions on the heavy compared to the light subbands of the 2DEG, the former coupling to phonons with frequencies ranging from below to above the bare band width. Our calculations reveal that electron-phonon coupling essentially results in an overall bandwidth renormalisation of these states, in agreement with our experimental data where we find the band bottom of the heavy state substantially above the value predicted by our model tight binding calculations. 
\begin{figure}[!h]
\begin{center}
\includegraphics[width=0.5\textwidth]{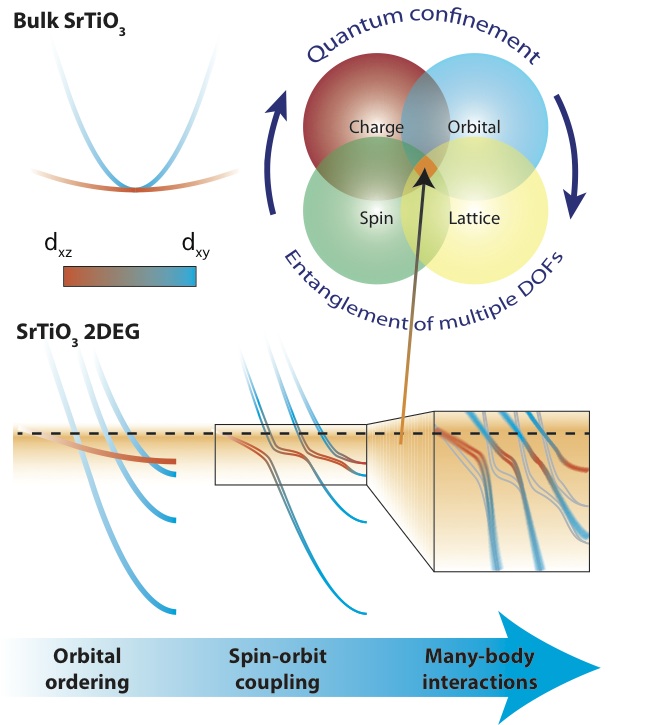}
\caption{ \label{f:summary} {\bf  Hierarchy and interplay of underlying degrees of freedom.} In the formation of the 2DEG, quantum confinement reconstructs the bulk electronic structure into a rich array of intersecting orbitally-ordered subband states. The interplay of this with spin-orbit coupling lifts the spin degeneracy of these bands, particularly strong around their crossings, through an orbitally-enhanced Rashba-like interaction. Electron-phonon and electron-electron interactions further renormalise these spin-split dispersions, increasing the quasiparticle masses close to the Fermi level, and causing a complete bandwidth narrowing of the shallow $d_{xz/yz}$ states that pins the orbitally-ordered band crossings close to the energy of the lowest phonon mode. This, in turn, sets the energy scale for enhanced spin-orbit splittings around these crossings, strongly entangling the charge, spin, orbital, and lattice degrees of freedom of the 2DEG on a low energy scale (orange shading).}
\end{center}
\end{figure}
\\
\
\\
{\large\noindent{\bf Discussion}}\\
The combination of electron-phonon coupling with orbital ordering therefore effectively pins the crossings of the $d_{xy}$ and $d_{xz/yz}$ subbands to the low-energy peak in the phonon density of states. As demonstrated above, however, we additionally find orbitally-enhanced spin-orbit splittings which become maximal around these band crossings. Our direct spectroscopic measurements, together with our theoretical calculations, therefore demonstrate how a co-operative effect of orbital ordering and electron-phonon coupling sets the relevant energy scale for dominant spin splitting in this system. 

Together, this reveals an intricate hierarchy of interactions and orderings governing the low-energy electronic structure of the SrTiO$_3$ 2DEG (Figure~\ref{f:summary}). 
Electrostatic screening in response to a surface or interface charge generates an electron accumulation layer confined to a narrow potential well. The resulting quantum size effects drive pronounced orbital ordering, creating multiple intersections of light $d_{xy}$- and heavy  $d_{xz/yz}$-derived subband states. This leads to orbital mixing and induces a significant local orbital angular momentum, which in turn permits a pronounced spin splitting to emerge, despite modest atomic spin-orbit coupling. Many-body interactions, of strikingly different form to the bulk, enhance the quasiparticle masses of these spin-split subbands, reduce their bandwidths, and renormalize the energetic locations of their intersections, thus modulating their unconventional spin splitting.

Together, this interplay strongly entangles the charge, spin, orbital, and lattice degrees of freedom of the SrTiO$_3$ 2DEG on an energy scale $\leq\!30$~meV. This will dominate both its transport and thermodynamic properties. Already we have shown how this explains enhanced quasiparticle masses observed from quantum oscillations as well as signatures of spin splitting in magnetotransport, unifying electrical and spectroscopic measurements from surface- and interface-based SrTiO$_3$ 2DEGs. More generally, it establishes how quantum size effects can dramatically manipulate the underlying electronic landscape of interacting electron liquids, setting the stage for engineering new emergent properties by dimensional confinement in transition-metal oxides.
\\
\
\\ 
{\small\noindent{\bf Methods}

\noindent{\it\textbf{Angle-resolved photoemission:}} ARPES measurements were performed at the CASIOPEE beamline of SOLEIL synchrotron, the SIS beamline of the Swiss Light Source, and beamline 10.0.1 of the Advanced light source using Scienta R4000 hemispherical electron analysers, and with base pressures below $5\times10^{-11}$~mbar. Single crystal SrTiO$_3$ commercial wafers were cleaved {\it in situ} at the measurement temperature of $T=20-30$~K along notches defining a $(100)$ plane. Measurements were performed on stoichiometric transparent insulating samples as well as very lightly La-doped samples (Sr$_{1-x}$La$_x$TiO$_3$ with $x=0.001$) to help avoid charging. 2DEGs were induced at the bare surface by exposure to intense UV synchrotron light.~\cite{Meevasana:NatureMater.:10(2011)114--118} The samples were exposed to irradiation doses $\gtrsim1000$~Jcm$^{-2}$ to saturate the formation of the 2DEG, and we experimentally confirmed that saturation was reached before starting any of the measurements presented here. The data shown here was measured using $s$-polarised photons of 51 or 55~eV, except for the Fermi surface maps shown in Supplementary Fig.~2 that used 43~eV $s$- and $p$-polarised light. All measurements were performed in the second Brillouin zone.
\\
\
\\
\noindent{\it\textbf{Self-energy determination:}} To determine the electron-phonon self-energy experimentally, we fit momentum distribution curves (MDCs) and energy distribution curves (EDCs) of the lowest $d_{xy}$ subband measured along the $[11]$ direction. We chose this band as its dispersion does not intersect that of the heavy $d_{xz/yz}$ subbands up to the chemical potential along this direction (Fig.~\ref{f:overview}(b)), allowing us to perform a quantitative analysis over an extended energy range, free from complications associated with the hybridization of different subbands. The real part of the self-energy, $\Sigma'(\omega)$, is given by the difference between our extracted dispersion and that of a `bare' band. In order to derive a Kramers-Kronig consistent self-energy, we take the cosine bare band shown in Fig.~\ref{f:e-ph}(a), which includes a moderate band width renormalization due to electron correlations. We extract the imaginary part of the self energy, $\Sigma''(\omega)=\Delta{k(\omega)}/2\cdot\partial\epsilon/\partial k$, where $\Delta{k}$ is the full width at half maximum of Lorentzian fits to MDCs, and $\partial\epsilon/\partial k$ is the bare band dispersion. This results in an imaginary part that includes a contribution from electron-electron interactions, which we approximate by the expression for a two-dimensional Fermi liquid $\Sigma_{\mathrm{e-e}}''(\omega)=\beta\omega^2[1+0.53 \ln(\omega/E_{\mathrm{F}})]$. Subtracting this contribution with realistic parameter values of $\beta=1.5$~eV$^{-1}$ and $E_{\mathrm{F}}=0.25$~eV from our total extracted $\Sigma''$ (see Fig.~\ref{f:e-ph}(b)) yields the imaginary part of the electron-phonon self-energy.
\\
\
\\
\noindent{\it\textbf{Electronic structure and self-energy calculations:}} To calculate the subband structure, we start from a relativistic density functional theory (DFT) calculation of bulk SrTiO$_3$ using the modified Becke-Johnson exchange potential and Perdew-Burke-Ernzerhof correlation functional  as implemented in the WIEN2K program~\cite{Blaha::()}. The muffin-tin radius of each atom $R_{\mathrm{MT}}$ was chosen such that its product with the maximum modulus of reciprocal vectors $K_{\mathrm{max}}$ become $R_{\mathrm{MT}}K_{\mathrm{max}}=7.0$. The Brillouin zone was sampled by a $15\times15\times15$ $k$-mesh. We downfold this using maximally-localized Wannier functions to generate a set of bulk tight-binding transfer integrals, and then incorporate these into a 30 unit cell supercell with additional on-site potential terms to account for band bending via an electrostatic potential variation. We solve this self-consistently with Poisson's equation, incorporating an electric field-dependent dielectric constant,~\cite{Copie:Phys.Rev.Lett.:102(2009)216804} to yield the bare band dispersions including Rashba-type spin splitting~\cite{Bahramy:NatCommun:3(2012)1159--} of the 2DEG. We stress that the total magnitude of the band bending is the only adjustable parameter, and yields a realistic electronic structure in good agreement with our spectroscopic measurements apart from our observed signatures of electron-phonon interactions which are not included at the level of DFT. To incorporate these, we calculate the self-energy $\Sigma_{\mathrm{e-ph}}(\omega)$ in an Eliashberg model, 
\begin{eqnarray*}
\Sigma_{\mathrm{e-ph}}(\omega) = \int_{-E_\mathrm{F}}^\infty N(\epsilon)d\epsilon\int_{0}^{\tilde{\omega}_{max}}d\tilde{\omega} \alpha^2 F(\tilde{\omega})\\
\times\bigg\{ \frac{f(-\epsilon,T)+n(\tilde{\omega},T)}{\omega-\epsilon-\tilde{\omega}+i\delta^{\pm}}+\frac{f(\epsilon,T)+n(\tilde{\omega},T)}{\omega-\epsilon+\tilde{\omega}+i\delta^{\pm}}\bigg\}
\end{eqnarray*}
where $N(\epsilon)$ is the bare electronic density of states determined from our tight binding calculations and $f(\epsilon,T)$ and $n(\tilde{\omega},T)$ are the Fermi and Bose occupation factors. For the coupling function $\alpha^2 F(\tilde{\omega})$ we use two different models. The blue line in Fig.~\ref{f:e-ph}(b) assumes $\alpha^2 F(\tilde{\omega})$ proportional to the entire O- and Ti-derived phonon density of states from Ref.~\cite{Choudhury:Phys.Rev.B:77(2008)134111}, while the black line is a calculation for the coupling strengths given in Refs.~\cite{Verbist:Ferroelectrics:130(1992)27--34,Meevasana:NewJ.Phys.:12(2010)023004} for the three longitudinal optical phonons that were found to dominate the electron-phonon interaction in lightly doped bulk samples.
We then calculate the spectral function
\[A(k,\omega)=-\frac{1}{\pi}\frac{\Sigma''(\omega)}{\left[\omega-\epsilon(k)-\Sigma'(\omega)\right]^2+\left[\Sigma''(\omega)\right]^2}\]
where the bare band dispersion $\epsilon(k)$ is taken from our tight-binding calculation. To better compare with our experimental data in Fig.~\ref{f:e-ph}(c), we project this onto different atomic orbitals, and include contributions from $d_{xy}$ and $d_{xz}$ but not $d_{yz}$ orbitals to account for transition matrix elements in our experimental geometry. We additionally project the calculation onto different layers of our supercell, and incorporate an exponential attenuation of signal with depth below the surface in photoemission, assuming an inelastic mean free path of 5~\AA, into our simulation. Finally, we convolve the simulated spectral function with a 2D Gaussian to account for an experimental energy and momentum resolution of 0.01~eV and 0.015~\AA$^{-1}$, respectively.}

\
\\
\
\noindent{\bf Acknowledgements:} This work was supported by the U.K. EPSRC (EP/I031014/1), the ERC (207901), the SNSF (200021-146995), the Scottish Funding Council, The Thailand Research Fund (RSA5680052), Office of the Higher Education Commission, Suranaree Univerisity of Technology, and the Japan Society for the promotion of Science (JSPS), through the `Funding Program for World-Leading Innovative R\&D on Science and Technology (FIRST Program)', initiated by the council for Science and Technology policy (CSTP). PDCK acknowledges support from the Royal Society through a University Research Fellowship (UF120096). We acknowledge SOLEIL (beamline CASSIOPEE), the ALS (beamline 10.0.1), and SLS (SIS beamline) for provision of synchrotron radiation facilities, and in particular N.C.~Plumb, M.~Radovi{\'c}, and M.~Shi (SLS) and P.~Le~F{\`e}vre, F.~Bertran and A.~Taleb-Ibrahimi (SOLEIL) for technical assistance. The Advanced Light Source is supported by the Director, Office of Science, Office of Basic Energy Sciences, of the U.S. Department of Energy under Contract No. DE-AC02-05CH11231. We gratefully acknowledge C.~Bell. C.~Berthod, V.~Cooper, A.~F\^ete, H.Y.~Hwang,  M.~Kim, J.~Mannhart, D.~van der Marel, and J.-M.~Triscone for useful discussions.

\
\\
\
\noindent{\bf Author contributions:} The experimental data was measured by PDCK, SMW, AT, ADLT, TE, PB, WM and FB, and analysed by PDCK, SMW, AT and FB. PDCK and MSB performed the electronic structure calculations and AT performed the electron-phonon self-energy calculations. SKM maintained the ARPES endstation at the Advanced Light Source and provided experimental support. PDCK and FB were responsible for overall project planning and direction, and wrote the manuscript with input and discussion from all co-authors.

\
\\
\
\noindent{\bf Competing financial interests:} The authors declare no competing financial interests.
\
\setcounter{figure}{0}
\renewcommand{\thefigure}{{S\arabic{figure}}}   
\
\begin{figure*}[!h]
\begin{center}
\includegraphics[width=0.7\textwidth]{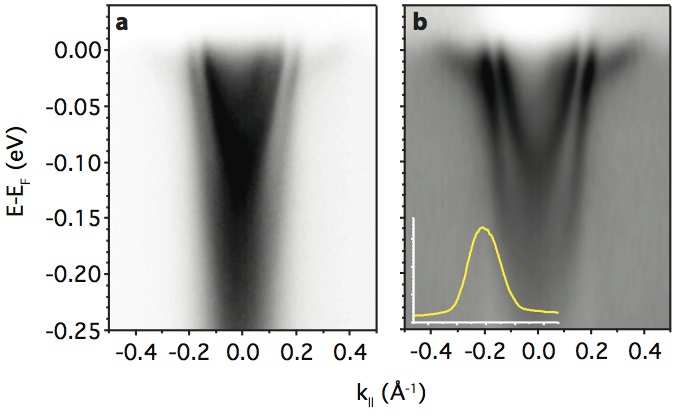}
\caption{ \label{fS:norm} {\bf Subband structure of the SrTiO$_3$ 2DEG.} (a) Raw data of the spectrum shown in Fig.~1a of the main manuscript, clearly revealing three light subband states as well as weak spectral weight from a single heavy band at lower binding energy. (b) This latter feature is strongly enhanced by dividing by the average MDC intensity (shown inset).}
\end{center}
\end{figure*}

\begin{figure*}[!h]
\begin{center}
\includegraphics[width=\textwidth]{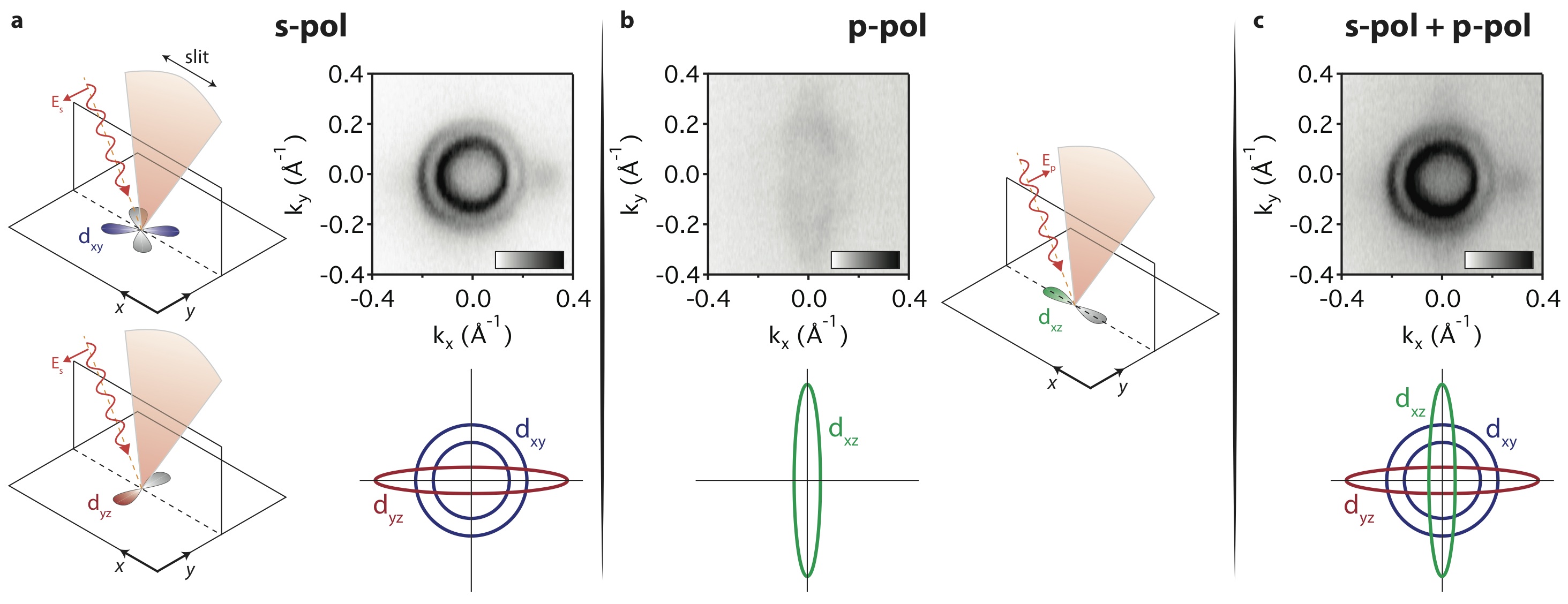}
\caption{ \label{fS:pol} {\bf Orbital character from polarisation-dependent ARPES.} Fermi surface of the SrTiO$_3$ 2DEG measured using (a) $s$-, (b) $p$-polarised, and (c) the sum of $s$- and $p$-polarised 43~eV photons, and normalised to the same total intensity scaling. A schematic of the electronic orbitals that should be dominant for the respective experimental geometries and the resulting Fermi surfaces are also shown, in good agreement with our assignment of dominant orbital characters of the subbands in the main text.}
\end{center}
\end{figure*}

\begin{figure*}[!h]
\begin{center}
\includegraphics[width=\textwidth]{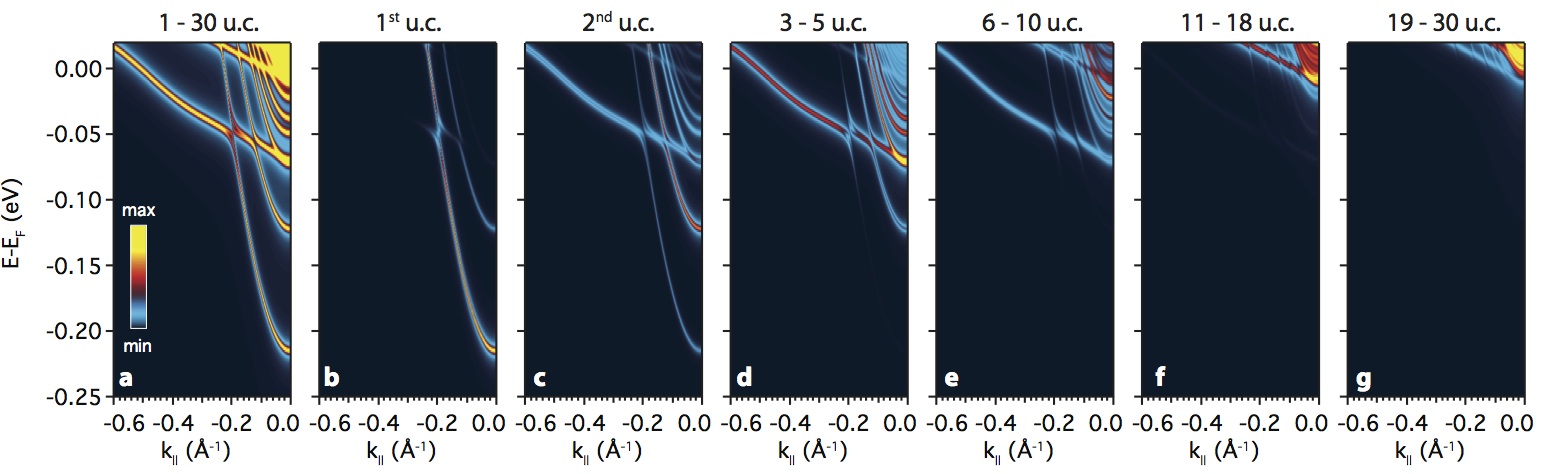}
\caption{ \label{fS:layer_proj} {\bf Spatial extent of the 2DEG.} Layer-projected calculations of the electronic structure of the SrTiO$_3$ 2DEG, integrated over (a) the full 30 unit cell (u.c.) supercell, and (b)--(g) individual or few unit cell regions, as labelled in the figure.}
\end{center}
\end{figure*}

\begin{figure*}[!h]
\begin{center}
\includegraphics[width=\textwidth]{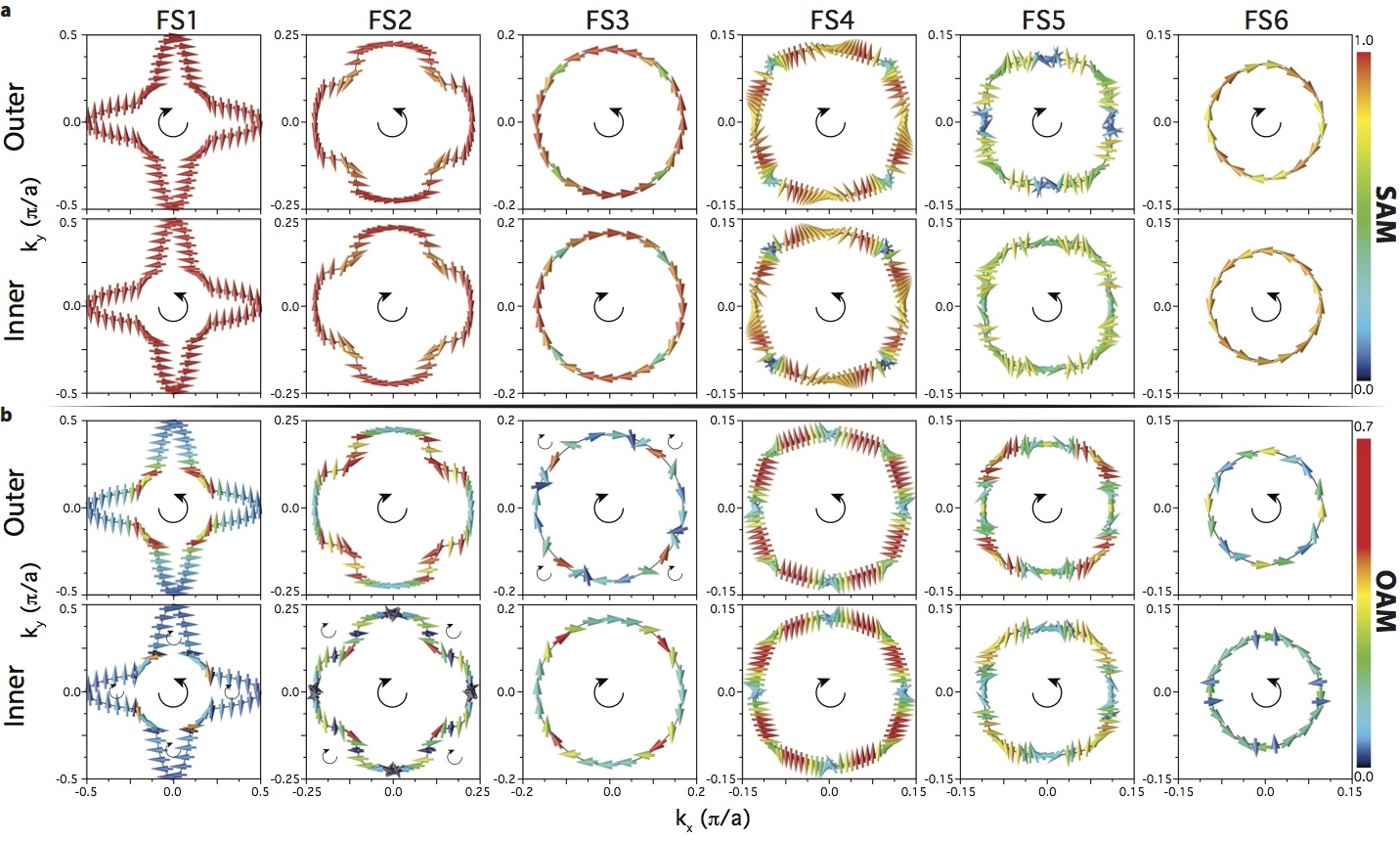}
\caption{ \label{fS:spin_text} {\bf Subband-dependent canted spin and orbital textures.} (a) In-plane spin texture (spin angular momentum, SAM) of the outer (top) and inner (bottom) branches of successive Fermi surface sheets, showing how the alignment of the spin direction evolves from radial- to tangential- as the orbital character changes from $d_{xy}$- to $d_{xz/yz}$-like (see Fig.~2(c) of the main text for the orbital character), leading to frustrated spin textures for orbitally-mixed Fermi surfaces. The net winding direction is shown in the centre of each plot, and the magnitude of the SAM is shown on a false colour scale in units of $\hbar/2$. (b) Equivalent plots for the orbital angular momentum (OAM), showing similar canting and mixed windings as represented by the large and small arrows. The magnitude of the OAM is represented by a false colour scale in units of $\hbar$, revealing how significant OAM emerges only around the band intersections where the orbital character becomes strongly mixed, and the spin splitting is thus maximised.}
\end{center}
\end{figure*}

\end{document}